# An alternative explanation of the 'spokes' observed in Saturn's Rings

Fenton J. Doolan (Independent Researcher)

## Abstract

Observed first by amateur astronomer Stephen J. O'Meara in the 1970s and then subsequently observed by the Voyager Spacecraft flybys in the early 1980s (Fig.1), it was realized that the 'spokes' flare out like spokes on a bicycle wheel [2]. Based on observations, it is evident that the behaviour of the 'spokes' in Saturn's rings is not primarily governed by gravitational interactions with the planet, moons, or ring material. In 2005, the Cassini probe provided confirmation that the 'spokes' are more likely influenced by Saturn's global magnetic field [20].

Here we show the 'spokes' that appear in Saturn's rings consist of grains of silicates coated in pyrolytic carbon through the process of Chemical Vapour Deposition (CVD). Pyrolytic carbon is a highly diamagnetic substance and can levitate above a sufficiently strong magnetic field. The 'spokes' also consist of ice particles that are diamagnetic as well. The photoelectric effect can be used to explain why the silicates coated in pyrolytic carbon return to the main ring structure when exposed to sunlight of a specific frequency. The pyrolytic carbon grains become paramagnetic when some of the unhybridised $2p_z$ orbitals lose their unpaired delocalised electrons, thus collapsing the π bond molecular orbital structure. The pyrolytic carbon grains are now attracted towards the magnetic field emanating above and below the main ring structure.

It is suggested that the 'spokes' in Saturn's B-ring are always present and that no plasma triggering event is required to increase plasma density. The 'spokes', however, are only visible when a favourable viewing angle is allowed, and their visibility is also dependent on the angle of the sunlight hitting Saturn's rings.

# 1 Introduction

Several models have been put forth to elucidate the formation of the 'spokes' observed in Saturn's rings. Among these models, the most widely accepted one by Morfill et al. (2003) proposes that the 'spokes' are generated as a result of meteorite impacts on the rings. These meteorite impacts produce a temporary cloud of dense plasma. This plasma, in turn, charges the dust particles, causing them to levitate above and below the plane of Saturn's B ring. According to this model, the 'spokes' are formed through resonate interactions between the oscillations within the rings and Saturn's rotating magnetosphere [14].

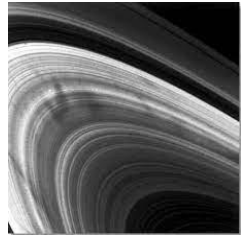

Figure 1: Voyager's image of the spokes (Image credit: JPL NASA)

An alternate model by Jones et al. (2006) suggests that the formation of the 'spokes' in Saturn's rings could be attributed to lightning-induced electron beams. According to this model, the electron beams impact specific regions of the rings that are magnetically connected to thunderstorms. The researchers propose that the density of Saturn's ionosphere plays a crucial role in determining where the 'spokes' are formed. The propagation of electron beams towards the rings may generate the observed X-ray emissions and introduce particles to Saturn radiation belts, consequently causing an alternation to the rings' composition over time [18].

According to research conducted by Goertz (1984), it is proposed that the formation of the 'spokes' in Saturn's rings is the result of electrostatically charged dust particles that are suspended in Saturn's magnetic field [13]. Unlike the ring particles that exhibit Keplerian motion around Saturn, these charged dust particles rotate in synchronization with the planet itself. However, when sunlight hits the B-ring at certain angles, these electrostatically charged dust particles lose their charge and settle back into the main ring structure causing the 'spokes' to disappear temporarily. The Cassini spacecraft observed that the 'spokes 'can form within minutes and fade away in a few hours, highlighting their transient nature [10].

## 2 Hypothesis

The small percentage of carbon that constitutes Saturn's rings may be diamagnetic pyrolytic carbon. During the formation of Saturn's protoplanetary disk, it is hypothesised, that pyrolytic carbon would have been deposited via Chemical Vapour Deposition (CVD) of hydrocarbon gases such as methane onto fine grains of silicates which acted as a substrate. These fine grains of silicates coated in pyrolytic carbon can levitate above or below a strong magnetic field due to pyrolytic carbon being highly diamagnetic. It is also suggested that Saturn's B-ring has a sufficiently strong magnetic field emanating orthogonally above and below its plane to levitate these pyrolytic carbon grains.

# 3 Justification of Hypothesis

In the laboratory it has been demonstrated that diamagnetic pyrolytic carbon levitates above a sufficiently strong magnetic field, see (Fig.2).

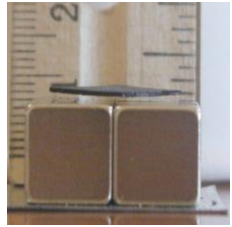

Figure 2: Levitating pyrolytic carbon (Image credit: scitoys.com)

Pyrolytic carbon is a man-made substance, but it is predicted that Saturn's rings consist of a small percentage of pyrolytic carbon. This type of carbon is formed in a vacuum at high temperatures of above 1400K, this process is known as flash vacuum pyrolysis. The dark 'spokes' which are observed in Saturn's B-ring consist of levitating particles that transition periodically from motion in synchrony with the rotation of Saturn's magnetic field to normal Keplerian motion within the main ring. The dark 'spokes' are only observed in Saturn's B-ring which corresponds to the 1500K region in the protoplanetary disk formation (Fig.3).

# 4 Research

According to research conducted by Henning et al. (2013), it is proposed that during Saturn's formation, the innermost regions of its protoplanetary disk would have reached these temperatures like those found in the 1500K region depicted in Figure 3. The specific temperature range corresponds to Saturn's B-ring, which explains why the dark 'spokes' are exclusively observed in this ring. Beyond the 1500K region, the temperatures would be too cold for 'spoke' formation to occur, hence their absence beyond Saturn's B-ring [16].

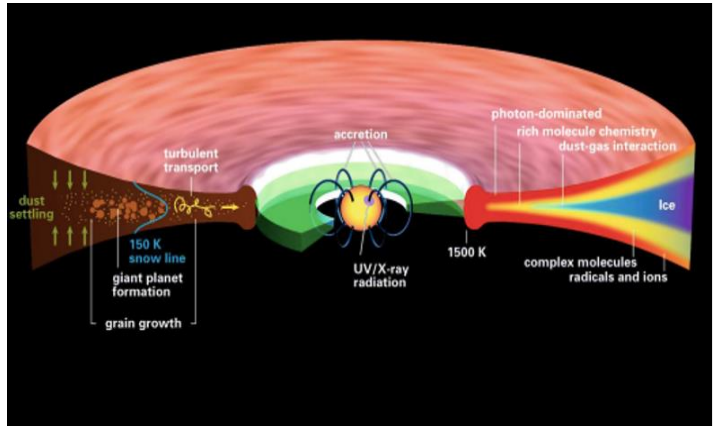

Figure 3: Protoplanetary disk formation (Image credit: astrochymist.org)

Saturn's ring system closely resembles the disc of dust and debris from which Earth and the other planets in the solar system originated approximately 4.55 billion years ago. This protoplanetary disc formed as result of gravitational collapse of a spherical cloud composed of extremely cold gas and dust. As the rotating cloud contracted, it transformed into a flat disc shape, exhibiting the swirling motion around the newly formed sun. "Once the sun had blown away the gas, the disc of orbiting rubble would have resembled the disc of Saturn's ring system" says Professor Carl Murray [4].

Protoplanetary disk surface temperatures, masses, and rate of mass in fall onto the disks can be estimated by observations of suspected planet-forming disks. The solar nebula's temperature is constrained to locations and times as suggested by the ongoing analysis of the formation of primitive meteorites, comets, and their components. Disk temperatures are in good agreement with theoretical models of disks undergoing accretion of mass due to an infalling cloud envelop. As such, the predicted temperatures on the inner disk are a moderately warm (500–1500K), surrounded by a cool (50–150K) outer disk [1].

When contemplating Saturn, one would instinctively associate the planet with extremely cold temperatures, making it seem unlikely for the required temperature of 1500K to be present for the formation of pyrolytic carbon. However, recent data presented by Fukuhara (2020) indicates that Saturn's core is actually very hot. The molten rocky metallic core of Saturn is estimated to have a temperature of 12200 $^0$C [12]. Thus, making Saturn's core remarkably hotter than what is commonly expected when thinking about the planet.

## Saturn's 'Ring Rain' chemical composition

Hydrocarbons such as methane can be converted to pyrolytic carbon at temperatures above 1400K as indicated in equation (1) below. Research by Spilker (2019) also indicates that the Cassini mission found an abundance of various hydrocarbons in the 'rain' produced by Saturn's rings (Fig. 4) which represents the composition of 'ring rain' produced by Saturn's rings [26]. It should be noted that silicates were also detected in the 'ring rain' which (Fig.4) neglects to show.

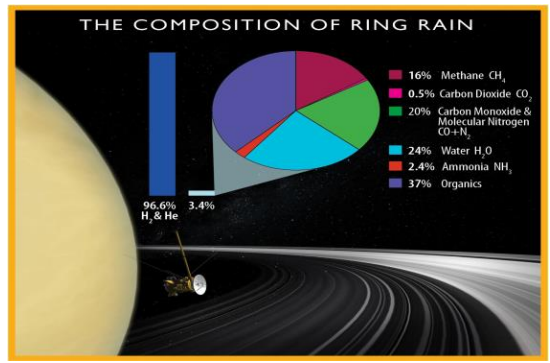

Figure 4: Composition of ring rain (Image credit: NASA/JPL/SwRI)

$$CH_{4(g)} + 1400K. \rightarrow C_{(s)} + 2H_{2(g)} \qquad (1)$$

During the investigation, the researchers made surprising discoveries regarding the composition of Saturn's atmosphere. They identified methane, ammonia, carbon monoxide, molecular nitrogen, and carbon dioxide which were all present to some extent.

However, the presence of methane and carbon dioxide was particularly unexpected. Contrary to their expectations, the researchers found a smaller amount of water ice. Molecular hydrogen was the most abundant constituent at all altitudes examined. Notably, observations revealed the presence of water falling from the rings, along with significant qualities of methane, ammonia, molecular nitrogen, carbon monoxide, carbon dioxide, and impact fragments of organic nanoparticles [31]. The in fall from the rings included substantial. amounts of water and hydrocarbons such as butane and propane [21].

## Pyrolysis and Gasification

Syngas, also known as synthesis gas, is a mixture of molecules that consist of hydrogen, methane, carbon monoxide, carbon dioxide, water vapour, along with other hydrocarbons and condensable compounds. It serves as the primary product of gasification and is predominantly generated through high temperature pyrolysis applied to various biomass, residues, and waste. During pyrolysis, volatile compounds within the raw material are vapourised by heat, triggering a sequence of intricate reactions [32]. Gases from pyrolysis typically contain large amounts of methane, hydrogen, carbon monoxide, and carbon dioxide, as well as larger hydrocarbons [25].

Gasification is a process that converts biomass into gases, yielding substantial amounts of nitrogen, carbon monoxide, hydrogen, and carbon dioxide. This conversion takes place at elevated temperatures (usually exceeding $700^0$C) without combustion, through precise control of oxygen/and or steam levels within the reaction. The resulting syngas is a highly combustible fuel due to its substantial water and carbon monoxide content. Further reactions occur between the carbon monoxide and residual water present in the organic material, resulting in the formation of methane and excess carbon dioxide, equation (2) [15].

$$4\,CO_{(g)} + 2\,H_2O_{(l)} \rightarrow CH_{4(g)} + 3\,CO_{2(g)} \qquad (2)$$

In gasification, reforming is a means to enhance the proportion of hydrogen by decomposing the hydrocarbons into carbon monoxide and hydrogen, equation (3). If the biomass is already made of hydrocarbons, reforming is the first stage towards syngas.

$$CH_{4(g)} + H_2O_{(l)} \rightarrow CO_{(g)} + 3\,H_{2(g)} \qquad (3)$$

The chemical processes of pyrolysis and gasification may explain where the ice and other constituents of Saturn's rings originated from. Thus, Robin Canup's hypothesis that Saturn's rings were formed when a Titan-sized moon with a rocky core and an icy mantle spiralled into Saturn may not be required [3].

## Diamagnetism

The diamagnetic properties of a material are determined by a property called magnetic susceptibility, $\chi$. The relationship between magnetic susceptibility and magnetic permeability is expressed in equation (4). A diamagnetic material has a magnetic susceptibility of less than zero and a minimum value of -1:

$$\chi_v = \mu_v - 1 \qquad (4)$$

$\mu_v$ is the magnetic permeability of the material. Superconductors are the most diamagnetic material (-1.04 x $10^{-3}$) followed by pyrolytic carbon (-4.09 x $10^{-4}$) then bismuth (-1.66 x $10^{-4}$) see Table 1.

| Materials | Diamagnetic strength x $10^{-5}$ SI units |
|---|---|
| Superconductor | -105 |
| Pyrolytic carbon | -40.9 |
| Bismuth | -16.6 |
| Mercury | -2.9 |
| Silver | -2.6 |
| Carbon(diamond) | -2.1 |
| Lead | -1.8 |
| Carbon(graphite) | -1.6 |
| Copper | -1.0 |
| Water | -0.91 |

Table 1: Diamagnetic materials (Image credit:byjus.com)

Diamagnetic forces induced in materials by a magnetic field have a very different behaviour from inverse-square law forces. The diamagnetic force depends on the gradient of the squared magnetic field as shown in equation (5):

$$\vec{F}_d = \chi^V \vec{\nabla} B^2 / 2\mu_o \qquad (5)$$

where, V is the volume of the diamagnetic material, $\mu_o$ is the vacuum magnetic permeability and B is the magnetic field [24].

Professor Vladimir Tchernyi and Sergei Kapranov (2023) have recently put forward a new mechanism known as the Tchernyi-Kapranov effect. This effect emphasizes the importance of the magnetic field of Saturn in the origin and evolution of Saturn's dense rings via the process of anisotropic accretion of diamagnetic ice particles during the formation of Saturn's protoplanetary cloud [27] [28].

## Magnetic susceptibility of pyrolytic carbon

Pyrolytic carbon which is highly diamagnetic may explain the dark 'spokes' observed in Saturn's rings. Because pyrolytic carbon is so diamagnetic (repelled by a magnetic field) none of it would have been detected in the 'ring rain' falling from Saturn's rings to its atmosphere. It is suggested that the dark 'spokes' are fine grains of silicates that have been covered in pyrolytic carbon due to the process of flash vacuum pyrolysis during the formation of Saturn. Pyrolytic carbon is the best-known material that displays the most similar properties to superconducting materials i.e., diamagnetism.

Strong variations in ion density in Saturn's electrically charged atmosphere produce strong coupling in the visible rings which consists mainly of electrically charged ice particles [30]. Saturn's rings are therefore electrically charged and thus must produce an electric field. Mitchell et al.(2006) state: "Because of the charging of the ring and the resulting electric field, the electron and ion densities immediately above the ring will not be equal" [20]. Since Saturn's rings are electrically charged with an associated electric field, it is proposed that the rings must emanate a magnetic field orthogonally above and below the ring plane.

As such, Saturn's rings can be considered an electromagnetic phenomenon, as also suggested by Professor Vladimir Tchernyi [29]. The ring's magnetic fields enable the fine grains of silicates covered in pyrolytic carbon to levitate above and below the main ice-ring structures due to their highly diamagnetic nature, thus producing the observed dark and bright 'spoke' structures that rotate in synchronization with the rotation of Saturn's magnetic field. Analysis of the spectral radiation power of the spokes provides a specific periodicity of about (640.6 +/- 3.5 min) which almost coincides with the period of rotation of the magnetic field of Saturn (639.4 min). Professor Tchernyi states: "Moreover, a strong correlation between the maxima and minima of activity of the spokes with the spectral magnetic longitudes is connected to the presence or absence of the radiation of Saturn's Kilometric Radiation (SKR)" [28].

The dark 'spokes' are most visible at the two seasonal equinoxes as the illumination of the rings is greatly reduced making possible unique observations highlighting features that depart from the ring plane i.e., the levitated fine grains of silicates covered in pyrolytic carbon.

The dark 'spokes' become visible mainly due to two reasons. The first is due to planet shine from Saturn which reflects off the ice particles in the rings but is absorbed by the fine grains of silicates covered in pyrolytic carbon during the seasonal equinoxes. Hedman et al. (2017) states: "We will highlight the importance of including illumination sources other than the Sun in the radiative transfer analysis, namely the Saturn-shine and the ring-shine" [9].

Secondly, due to an increase in plasma density caused by the reduced illumination of Saturn's rings at the equinoxes, allowing the silicates coated in pyrolytic carbon to become 'recharged'. Thus, regaining their diamagnetic properties, enabling them to levitate above and below Saturn's B-ring plane.

In a study conducted by a research group in Japan in 2012, led by Professor Kobayashi, it was demonstrated that pyrolytic carbon has the ability to respond to laser light or sufficiently powerful natural sunlight by spinning or moving in the direction of the field gradient. When subjected to sufficient illumination the carbon's magnetic susceptibility weakens, leading to an unbalanced magnetization of the material. This phenomenon leads to movement when using a specific geometry [19].

This may explain why the dark 'spokes' in Saturn's rings appear seasonally at the equinoxes when the illumination from the sun is at a minimum. Hence the pyrolytic carbon grains levitate above and below the main ring. Illumination from the sun causes a weakening of the pyrolytic carbon's magnetic susceptibility, causing the pyrolytic carbon grains to return to the main ring.

## Hybridisation of Carbon

During the process of Chemical Vapour Deposition (CVD) of methane gas the carbon atoms would share the hydridised sp$^2$ electrons with their three neighbouring carbon atoms (Fig.5). Thus, forming a layer of honeycomb net-

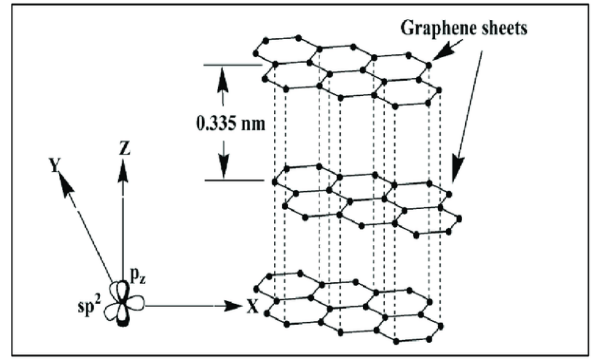

Figure 5: The formation of sp$^2$ hybrid orbitals [11]

work of planar structure (120$^0$ bond angle), which is also known as monolayer graphene. In pyrolytic carbon, these monolayers would form a turbostratic structure i.e. the graphene layers are arranged without order but have some covalent links between the layers (Fig.6)

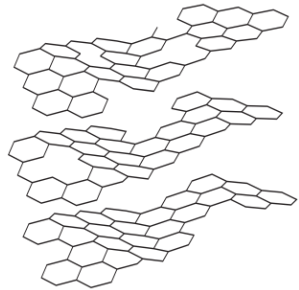

Figure 6: Turbostratic structure of graphene monolayer (Image credit: Wiki Common)

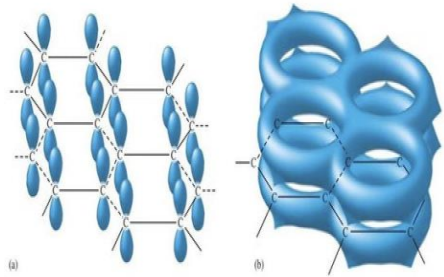

Figure 7: a. Valence bond theory $2p_z$ orbital depiction b. $\pi$ molecular orbital depiction (Image credit: Slide Serve)

When sunlight of a specific frequency hits the surface of the pyrolytic carbon grains the delocalised $\pi$ electrons in the molecular orbital (Fig. 7) created by the overlap of the unhybridised $2p_z$ orbitals collapse due to a decrease in electron density caused by electron ejection via the photoelectric effect.

According to valence bond theory, the unhybridised $2p_z$ orbitals will be vacant or will have an unpaired electron, thus making the pyrolytic carbon grains highly paramagnetic. The pyrolytic carbon grains are now attracted towards the magnetic field emanating from Saturn's B-ring. After returning to the main ring structure the pyrolytic carbon grains become 'recharged' when the plasma density reaches a maximum, this occurs when the sun's illumination of the rings is at a minimum i.e., at the equinoxes. The unhybridised $2p_z$ orbitals are now able to re-establish their $\pi$ molecular orbital structure due to the increase in electron density. Due to electromagnetic induction, the electrons in the $\pi$ bonds become highly diamagnetic.

## Electromagnetic Induction

The silicates coated in pyrolytic carbon become diamagnetic when exposed to the strong magnetic field emanating orthogonally from Saturn's B-ring plane.

Due to the process of electromagnetic induction, a changing magnetic flux is produced in the delocalised electrons in the pyrolytic carbon's π molecular orbitals. Faraday's Second Law of Electromagnetic Induction equation (6) stipulates that there will be an EMF induced in the electrons in the $2p_z$ unhybridized orbitals which will tend to oppose the changing magnetic flux according to Lenz's law, equation (7). Faraday's second law of electromagnetic induction requires that the induced EMF to be equal to the rate of change of flux linkage. Faraday's Second Law of Electromagnetic Induction states,

$$\varepsilon = N \frac{\Delta \Phi}{\Delta t} \tag{6}$$

hence Lenz's Law states,

$$\varepsilon = -N \frac{\Delta \Phi}{\Delta t} \tag{7}$$

where $\varepsilon$ is the induced emf, N is the number of spin-orbits, $\Delta \Phi$ is the change in magnetic flux and $\Delta t$ is the change in time.

Thus, there will be an EMF induced that tends to oppose the magnetic field. The EMF will induce a larger current in some of the $2p_z$ delocalised electron's spin-orbits compared to other $2p_z$ delocalised electrons which have opposing spin-orbits. The resulting effect is that the magnetic moments now do not cancel out. Thus, making all the $2p_z$ delocalised electrons in the pyrolytic carbon act like tiny magnets, consequently, the pyrolytic carbon grains will be highly diamagnetic and will be repelled by the magnetic field emanating above and below Saturn's B-ring.

# 5   Discussion

Only when the sun is at the equinoxes resulting in the illumination of the rings to be at a minimum, does the plasma density above and below Saturn's rings reach a maximum. As such no triggering mechanism such as lightning is required to increase the plasma density above or below the ring plane is required. Due to the increased plasma density the $2p_z$ unhybridised orbitals can gain electrons reestablishing the π molecular orbitals on the surface of the pyrolytic carbon. Through the process of electromagnetic induction, the pyrolytic carbon grains become highly diamagnetic causing them to levitate above and below Saturn's B-ring.

The dark 'spokes' in Saturn's B-ring become visible due to the backscattering of light (Fig.9). This proposed mechanism suggests that Saturn's B-ring produces a magnetic field that emanates orthogonally above and below its ring plane.

The bright 'spokes' in (Fig.8) are visible on the unilluminated side of Saturn's rings when the sun's illumination of the rings is at a maximum (directly below or above the ring plane). The bright 'spokes' can be explained due to the forward scattering of light (Fig.9) caused by the grains of pyrolytic carbon which levitate above the plane of Saturn's rings.

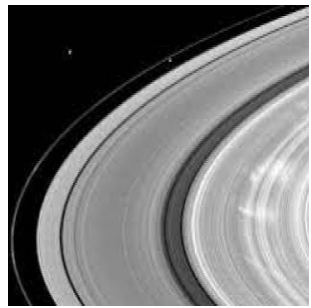

Figure 8: Bright spokes in Saturn's B-ring observed by Cassini (Image Credit: Smithsonian National Air and Space Museum)

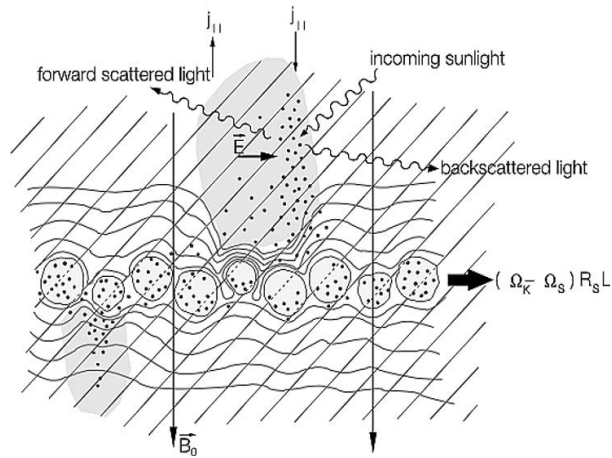

Figure 9: Forward and backscattering of light [17]

The bright 'spokes' which appear on the illuminated side of Saturn's rings may be due to the sunlight reflecting off the small, levitated ice particles. These small ice particles are also diamagnetic but unlike the silicates covered in pyrolytic carbon, they remain unaffected by natural sunlight as they are not magnetically susceptible to natural sunlight i.e. the photoelectric effect will not occur, so they remain levitated above Saturn's main rings.

Hence, the bright 'spokes' should always be observable given sufficient illumination by the sun and the correct angle of observation by the observer. The bright 'spokes' may suddenly disappear when the illumination of the sun reaches a critical point, causing the levitated ice particles to sublimate due to the sunlight's intensity.

## Cassini / VIMS Spectrometer

The first detection of 'spokes' in Saturn's B-ring was made in July 2018 using Cassini's VIMS spectrometer. A wide range of wavelengths (0.35-0.51 µm) was utilized to obtain the first complete reflectance spectrum of multiple 'spokes' (Fig.10).

The spectrum analysis confirmed that the 'spokes' consist of spheroidal water ice particles. These ice particles exhibit a wide size distribution, with a modal radius of approximately 1.90 μm, which is significantly larger than previously modelled sizes. It has been verified by spectral analysis that the bright 'spokes' on the illuminated side of Saturn's rings are water ice particles. The spectral analysis of the rings showed that $H_2O$ dominates in the near-IR. However, there was a sharp decrease in reflectance below approximately 0.6 μm, indicating the presence of some other substance [8].

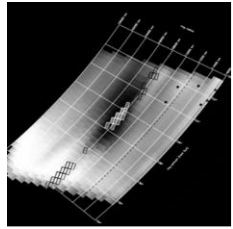

Figure 10: Cassini/VIMS spectral image of a spoke [8]

Recent studies by Ciarniello et.al, (2019) focused on the visible and near-IR spectra of Saturn's rings, identified reddish organic-rich refractory materials known as tholins as a potential contaminant responsible for the distinct reddish colour of the rings at visual wavelengths [5].

In a study by Cuzzi and Estrada (1998), the researchers investigated the colour of Saturn's rings and identified a material that imparts a distinct red colour to the particles in the A and B rings. They noted that "No silicates have the appropriate combination of steep spectral slope and high absorptivity to explain the rings' visual colour while remaining compatible with microwave observations." They found that incorporating Titan tholins into the water ice grains matched the colours and albedos of the particles. They also suggested that the darker rings may contain a "material with properties like carbon black, as seen in at least some comets and interplanetary dust particles, is needed" [7].

Modelling by Poulet and Cuzzi (2002) and Poulet et al. (2003) suggest a mixture of tholins and amorphous carbon to achieve fits to the observational data in the 0.3–4.0 µm range [23]. Poulet et al. (2003), concluded further that, for the A and B rings, while the carbon grains are intimately mixed in a "salt and pepper" fashion with the ice, the tholins are mixed at the molecular level within the ice grains themselves [22].

According to Cruikshank et al. (2005), tholins exhibit relatively few and weak absorption features in the VIMS near-IR region, except at wavelengths near 3.0 and 4.5 µm where water ice is also exhibits high absorption [6].

No dark 'spokes' will be observable on the underneath side of Saturn's B-ring due to the illumination from the sun causing the pyrolytic carbon grains to move back to the main ring due to the photoelectric effect. The pyrolytic carbon grains lose some of their delocalised unhybridised $2p_z$ electrons thus becoming paramagnetic. Now the pyrolytic carbon grains can move back to the main ring due to being attracted to the magnetic field. The dark 'spokes' become visible only at the equinoxes due to the sun's minimum level of illumination of the rings and backscattering of light. The plasma density above and below Saturn's B-ring increases to a maximum causing the pyrolytic carbon $2p_z$ orbitals to be 'recharged' i.e. gain electrons. Thus, the pyrolytic carbon grains can reform their π molecular orbitals and regain their highly diamagnetic nature, enabling them to levitate above and below Saturn's B-ring plane.

# 6   Conclusion

The chemical process of flash vacuum pyrolysis has converted hydrocarbons such as methane to pyrolytic carbon at temperatures above 1400K during the protoplanetary disk formation of Saturn. The pyrolytic carbon has coated the fine grains of silicates through the process of Chemical Vapour Deposition (CVD). The silicates coated in pyrolytic carbon are now able to levitate above or below the magnetic field emanating from Saturn's rings, due to the highly diamagnetic nature of pyrolytic carbon. If some of the 'spokes' do consist of pyrolytic carbon, then this would be proof that Saturn's rings were formed after the collapse of the protoplanetary cloud during the formation of Saturn. Thus, the argument concerning the age of Saturn's rings would therefore be put to rest once and for good.

Depending on the angle and frequency of the sunlight hitting the fine grains of silicates coated in pyrolytic carbon the photoelectric effect will cause the ejection of some of the delocalised π electrons from the pyrolytic carbon's molecular orbitals. This causes the pyrolytic carbon's molecular orbitals to collapse into discrete unpaired $2p_z$ unhybridised orbitals, thus becoming highly paramagnetic. The pyrolytic carbon grains will now return to the main ring as they are attracted to the magnetic field emanating from Saturn's B-ring.

The dark 'spokes' in Saturn's B-ring are only observable at the equinoxes due to the minimum illumination of its rings by the sun. The bright 'spokes' should be visible when the sun is above or below the plane of Saturn's rings. Therefore, it can be concluded that Saturn always has 'spokes', but their type either dark or bright depends on the position of the sun relative to Saturn and the angle at which the observer is observing the rings.

Saturn's rings are an electromagnetic phenomenon as suggested by Russian Professor Vladimir Tchernyi (2021) [29]. Saturn creates electromagnetic fields which encompass the equatorial region of the planet. These electromagnetic fields consist of magnetic fields which emanate orthogonally above and below the ring plane. Saturn's rings are analogous to magnetic field lines produced in a laboratory using a neodymium magnet and iron filings as shown in (Fig.11).

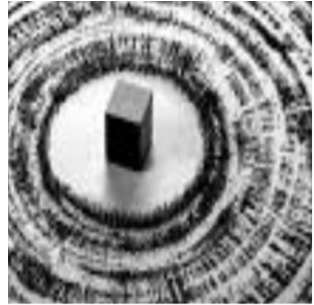

Figure 11: The Magnetic field lines created by a neodymium magnet look similar to Saturn's rings: (Image credit: Phys.org )

It is also predicted that the recent samples taken from the carbonaceous asteroids Bennu and Ryugu will contain a large quantity of pyrolytic 'diamagnetic' carbon.

Further research is suggested concerning the processes of pyrolysis and gasification as possible chemical processes that may help to explain where all Earth's water originated from.

# References

[1] Alan P Boss. Temperatures in protoplanetary disks. *Annual Review of Earth and Planetary Sciences*, 26(1):53–80, 1998.

[2] James Bryan. Stephen J. O'Meara and ring spokes before Voyager 1. *Journal of Astronomical History and Heritage*, 10:148–150, 2007.

[3] Robin M Canup. Origin of Saturn's rings and inner moons by mass removal from a lost Titan-sized satellite. *Nature*, 468(7326):943–946, 2010.

[4] Marcus Chown. Cassini: Lord of the Rings. *New Scientist*, 208(2789):40-41, 2010.

[5] Mea Ciarniello, Gianrico Filacchione, Emiliano D'Aversa, Fabrizio Capaccioni, PD Nicholson, JN Cuzzi, RN Clark, MM Hedman, CM Dalle Ore, Priscilla Cerroni, et al. Cassini-vims observations of Saturn's main rings: Ii. a Spectrophotometric study by means of Monte Carlo Ray-Tracing and Hapke's theory. *Icarus*, 317:242–265, 2019.

[6] Dale P Cruikshank, Hiroshi Imanaka, and Cristina M Dalle Ore. Tholins as coloring agents on outer Solar System bodies. *Advances in Space Research*, 36(2):178–183, 2005.

[7] Jeffrey N Cuzzi and Paul R Estrada. Compositional evolution of Saturn's rings due to meteoroid bombardment. *Icarus*, 132(1):1–35, 1998.

[8] E D'Aversa, G Bellucci, F Altieri, FG Carrozzo, G Filacchione, F Tosi, PD Nicholson, MM Hedman, RH Brown, and MR Showalter. Spectral characteristics of a spoke on the Saturn rings. *Memorie della Societa Astronomica Italiana Supplementi*, 16:70, 2011.

[9] Emiliano D'Aversa, Giancarlo Bellucci, Gianrico Filacchione, Priscilla Cerroni, Phil D Nicholson, Filippo G Carrozzo, Francesca Altieri, Fabrizio

Oliva, Anna Geminale, Giuseppe Sindoni, et al. IR spectra of Saturn's ring spokes and multiple shines in the Saturn-rings system. In *EGU General Assembly Conference Abstracts*, page 9140, 2017.

[10] AJ Farmer and P Goldreich. Spokes in Saturn's rings. In *American Astronomical Society Meeting Abstracts*, volume 205, pages 43–07, 2004.

[11] Furqan Farooq, Arslan Akbar, and R Arsalan Khushnood. Effect of hybrid carbon nanotubes/graphite nano platelets on mechanical properties of cementitious composite. In *Proceedings of the 1st Conference on Sustainability in Civil Engineering*, 2019.

[12] Mikio Fukuhara. Possible nuclear fusion of deuteron in the cores of Earth, Jupiter, Saturn, and brown dwarfs. *AIP Advances*, 10(3):035126, 2020.

[13] CK Goertz. Formation of Saturn's spokes. *Advances in Space Research*, 4(9):137–141, 1984.

[14] TW Hartquist, O Havnes, and GE Morfill. The effects of charged dust on Saturn's rings. *Astronomy & Geophysics*, 44(5):5–26, 2003.

[15] Steffen Heidenreich and Pier Ugo Foscolo. New concepts in biomass gasification. *Progress in energy and combustion science*, 46:72–95, 2015.

[16] Thomas Henning and Dmitry Semenov. Chemistry in protoplanetary disks. *Chemical Reviews*, 113(12):9016–9042, 2013.

[17] M Horányi, TW Hartquist, O Havnes, DA Mendis, and GE Morfill. Dusty plasma effects in Saturn's magnetosphere. *Reviews of Geophysics*, 42(4), 2004.

[18] GH Jones, N Krupp, H Krüger, E Roussos, W-H Ip, DG Mitchell, SM Krimigis, J Woch, A Lagg, M Fränz, et al. Formation of Saturn's ring spokes22

by lightning-induced electron beams. *Geophysical research letters*, 33(21), 2006.

[19] Masayuki Kobayashi and Jiro Abe. Optical motion control of Maglev graphite. *Journal of the American Chemical Society*, 134(51):20593–20596, 2012.

[20] CJ Mitchell, M Horányi, O Havnes, and CC Porco. Saturn's spokes: Lost and Found. *Science*, 311(5767):1587–1589, 2006.

[21] Mark Perry, Hunter Waite, Rebecca Perryman, Don Mitchell, Tom Cravens, Luke Moore, Kelly Miller, Roger Yelle, Ben Teolis, and Ralph McNutt. The flow of material inward from Saturn's rings. In *EGU General Assembly Conference Abstracts*, page 16834, 2018.

[22] F Poulet, DP Cruikshank, JN Cuzzi, TL Roush, and RG French. Compositions of Saturn's rings a, b, and c from high resolution near-infrared spectroscopic observations. *Astronomy & Astrophysics*, 412(1):305–316, 2003.

[23] F Poulet and JN Cuzzi. The composition of Saturn's rings. *Icarus*, 160(2):350–358, 2002.

[24] Guilherme Reis. Thermal influence on the diamagnetic properties of pyrolytic graphite: Applications in space and high-speed transportation. 2016.

[25] Jens Rostrup-Nielsen and Lars J Christiansen. *Concepts in syngas manufacture*, volume 10. World Scientific, 2011.

[26] Linda Spilker. Cassini-Huygens' exploration of the Saturn system: 13 years of discovery. *Science*, 364(6445):1046–1051, 2019.

[27] Vladimir V Tchernyi and Sergey V Kapranov. How Saturn could create dense rings after the emergence of its magnetic field. The Tchernyi-Kapranov effect: Mechanism of magnetic

anisotropic accretion. *Physics & Astronomy International Journal*, 7(1):54–57, 2023.

[28] Vladimir V Tchernyi and Sergey V Kapranov. Contribution of magnetism to the origin of Saturn's rings. *The Astrophysical Journal*, 894(1), 2020.

[29] Vladimir V Tchernyi, Sergey V Kapranov, and Andrey Yu Pospelov. Role of electromagnetism in the origin of Saturn's rings due to diamagnetism of their ice particles: JC Maxwell had almost solved the rings origin problem. *URSI Radio Science Letters*, 3:69, 2021.

[30] JE Wahlund, MW Morooka, L Hadid, DJ Andrews, WS Kurth, G Hospo-darsky, and AM Persoon. First in-situ determination of the ionospheric structure of Saturn by Cassini/rpws. *Saturn*, 2017.

[31] JH Waite Jr, RS Perryman, ME Perry, KE Miller, J Bell, TE Cravens, CR Glein, J Grimes, M Hedman, J Cuzzi, et al. Chemical interactions between Saturn's atmosphere and its rings. *Science*, 362(6410):eaat2382, 2018.

[32] Biogreen https://www.biogreen-energy.com/syngas

**The author declares there are no conflicts of interest regarding this article.**